\documentstyle[epsfig]{mn2e}
\begin{document}
\input BoxedEPS.tex
\newcommand{\aip}{{\small ${\cal AIPS}$}}
\newcommand{\gtsim}{\mbox{{\raisebox{-0.4ex}{$\stackrel{>}{{\scriptstyle\sim}}
$}}}}
\newcommand{\ltsim}{\mbox{{\raisebox{-0.4ex}{$\stackrel{<}{{\scriptstyle\sim}}
$}}}}
\newcommand{\s}{$\stackrel{\rm s}{.}$}
\newcommand{\h}{$^{\rm h}$}
\newcommand{\m}{$^{\rm m}$}
\newcommand{\pp}{$\stackrel{\prime\prime}{.}$}
\newcommand{\de}{$^{\circ}$}
\newcommand{\p}{$^{\prime}$}
\newcommand{\arc}{$^{\prime\prime}$}
\newcommand{\marc}{^{\prime\prime}}
\newcommand{\rs}{{\em $r_s$}}
\newcommand{\DPM}{{\em DPM}}
\newcommand{\alf}{{\displaystyle\biggl({\nu_{\rm h} \over \nu_{\rm l}}\biggr)^{\alpha}} }

\newcommand{\figstart}[1]
    { \begin{figure}[htb]
      \begin{picture}(0,#1) }
\newcommand{\figend}[4]
    { \end{picture}
      \special{#1}
      \caption[#2]{#3}
      \label{#4}
      \end{figure} }
\newcommand{\fig}[5]
    { \figstart{#1}
      \figend{#2}{#3}{#4}{#5} }
\newcommand{\bHS}{\beta_{\mbox{\scriptsize HS}}}
\newcommand{\bBF}{\beta_{\mbox{\scriptsize BF}}}
\newcommand{\nT}{\nu_{\mbox{\scriptsize T}}}
\newcommand{\et}{E_{\mbox{\scriptsize T}}}
\newcommand{\nTn}{\nu_{\mbox{\scriptsize Tn}}}
\newcommand{\nTf}{\nu_{\mbox{\scriptsize Tf}}}
\newcommand{\tn}{\tau_{x\mbox{\scriptsize n}}}
\newcommand{\tf}{\tau_{x\mbox{\scriptsize f}}}
\newcommand{\xn}{x_{\mbox{\scriptsize n}}}
\newcommand{\xf}{x_{\mbox{\scriptsize f}}}
\newcommand{\yn}{y_{\mbox{\scriptsize n}}}
\newcommand{\yf}{y_{\mbox{\scriptsize f}}}
\newcommand{\lln}{l_{\mbox{\scriptsize n}}}
\newcommand{\llf}{l_{\mbox{\scriptsize f}}}
\newcommand{\Dn}{f(\Delta_{\mbox{\scriptsize n}})}
\newcommand{\Df}{f(\Delta_{\mbox{\scriptsize f}})}
\newcommand{\B}{\mbox{$B$}}
\newcommand{\Bo}{\mbox{$B$}_{0}}

\SetEPSFDirectory{/scratch/sbgs/figures/hst/}
\SetRokickiEPSFSpecial
\HideDisplacementBoxes

\title[Modelling the Spoon IRS diagnostic diagram]{Modelling the Spoon IRS diagnostic diagram}
\author[Rowan-Robinson M.]{Michael Rowan-Robinson$^1$, Andreas Efstathiou$^{2}$\\
$^1$ Astrophysics Group, Blackett Laboratory, Imperial College of Science 
Technology and Medicine, Prince Consort Road,London SW7 2AZ\\
$^2$  School of Sciences, European University Cyprus, Diogenes St, Engomi, 1516 Nicosia, Cyprus.\\ 
}
\maketitle
\begin{abstract}
We explore whether our models for starbursts, quiescent star-forming galaxies and for AGN dust tori are able to 
model the full range of IRS spectra measured with {\it Spitzer}.  The diagnostic plot of 9.7 $\mu$m silicate
optical depth versus 6.2 $\mu$m PAH equivalent width, introduced by Spoon and coworkers in 2007, gives a good indication of the age
and optical depth of a starburst, and of the contribution of an AGN dust torus.   However there is aliasing
between age and optical depth at later times in the evolution of a starburst, and between age 
and the presence of an AGN dust torus. 

Modeling the full IRS spectra and using broad-band 25-850 $\mu$m fluxes can help to resolve these aliases.  
The observed spectral energy distributions require starbursts of a range of ages with initial dust optical depth
ranging from 50-200, optically thin dust emission ('cirrus') illuminated by a range of surface brightnesses of the
interstellar radiation field, and AGN dust tori with a range of viewing angles.

\end{abstract}


\section{Introduction}
The wealth of mid and far infrared spectroscopy of galaxies from {\it ISO} and {\it Spitzer} have provided a challenge
for models of their spectral energy distributions (SEDs), which has been tackled by several groups. 
Over the past twenty years we have
developed increasingly sophisticated radiative transfer models for different types of infrared galaxy,
for example for  starburst galaxies
(Rowan-Robinson \& Crawford 1989, Rowan-Robinson \& Efstathiou 1993, Efstathiou et al 2000),
AGN dust tori (Rowan-Robinson \& Crawford 1989, Efstathiou \& Rowan-Robinson 1995, Rowan-Robinson 1995), and quiescent
 ('cirrus') galaxies (Rowan-Robinson 1992, Efstathiou \&Rowan-Robinson 2003, Efstathiou \& Siebenmorgen 2009).  
Each of these model types involves at least two significant model parameters so there are a great 
wealth of possible models, particularly as a galaxy SED may be a mixture of all three types.
Starburst models have also been developed by Silva et
al. (1998), Takagi et al. (2003), Dopita et al.  (2005), Siebenmorgen \& Krugel (2007).
Other work on radiative transfer modeling of the torus in AGN
has been presented by Pier \& Krolik (1992), Granato \& Danese (1994),
Nenkova et al. (2002, 2008), Dullemond \& van Bemmel (2005), 
H\"{o}nig et al. (2006) and Schartmann et al. (2008). Other work on
radiative transfer modeling of cirrus galaxies has been presented by
Silva et al (1998), Dale et al (2001) and Piovan et al (2006). 

Often, however, we have only limited broad-band data available and in this situation it is more illuminating
to use a small number of infrared templates to match the observed infrared colours (eg Rowan-Robinson
and Crawford 1989, Rowan-Robinson 1992, 2001, Rowan-Robinson and Efstathiou 1993,
Rowan-Robinson et al 2004, 2005, 2008).   These templates 
have proved remarkably successful in matching observed Spitzer SEDs, including cases where IRS data are 
available (Rowan-Robinson et al 2006, Farrah et al 2008, Hernan-Caballero et al 2009).

Spoon et al (2007) have published a very interesting diagnostic diagram for starburst and active galaxies, which
plots the strength of the silicate 9.7 $\mu$ feature against the equivalent width (EW) of the 6.2 $\mu$m PAH feature
for 180 galaxies with Spitzer IRS spectra.  In this paper we explore how well our models fit the distribution 
of galaxies in this diagram.

\section{Description of the models}

\subsection{AGN torus models}

 The most important constraint on early
models for the torus in AGN (Pier \& Krolik 1992,
Granato \& Danese 1994, Efstathiou \& Rowan-Robinson 1995)
was provided by mid-infrared spectroscopic observations from the ground
in the $8-13\mu m$ window by Roche, Aitken and collaborators
(Roche et al 1991). The observations showed moderate
absorption features in type 2 AGN and featureless spectra
in type 1 AGN. By comparison the flared discs  
(whose thickness increases linearly with distance from the central source)
of Efstathiou \& Rowan-Robinson (1990, 1995) and Granato \& Danese
(1994) showed strong silicate emission features when observed face-on. The cylindrical
geometry and high optical depth of the Pier \& Krolik models
eliminated the emission features but the overall spectra were
rather narrow. Granato \& Danese suggested that the silicate
grains are destroyed  by shocks in the inner part of the torus.
Efstathiou \& Rowan-Robinson proposed that tapered discs (whose
thickness increases with distance from the central source in
the inner part of the disc but tapers off to a constant value
 in the outer disc) with a density distribution that followed $r^{-1}$
could give flat spectra in the mid-infrared for an opening angle
of around $45^o$ and an equatorial optical depth at 1000$\AA$ of 1000.
 Models with a smaller opening angle predicted
face-on spectra with shallow silicate absorption features
whereas models with a larger opening angle showed weak emission
features. Recent {\em Spitzer} observations showed weak emission
features in quasars and weak absorption features in Seyfert 1 galaxies
(e.g. Hao et al 2005, 2007, Siebenmorgen et al 2005, Spoon et al 2007).
 The tapered discs of Efstathiou \& Rowan-Robinson, in combination with
 the starburst models of Efstathiou, Rowan-Robinson \& Siebenmorgen (2000),
 which are described below, have been very successful in
fitting the spectral energy distributions of a number of AGN
(e.g. Alexander et al 1999, Ruiz et al 2001, Farrah et al 2003,
 Efstathiou \& Siebenmorgen 2005) 
More recently there has been interest in clumpy torus models 
(Nenkova et al 2002, 2008, Dullemond \& van Bemmel 2005, 
H\"{o}nig et al 2006) which as suggested by Rowan-Robinson (1995)
also display weaker silicate features. 

In this paper we 
will use the tapered disc models of Efstathiou \& Rowan-Robinson (1995).
The models assume that the dust is smoothly distributed in the disc.
However, this distribution may be considered a reasonable approximation to a clumpy dust
distribution if the mean distance between clouds is comparable
to their size. For this paper we have explored a model in which we
fix the equatorial 1000$\AA$  optical depth at 1000, the opening angle of the torus
at 60$^0$, and the ratio of inner to outer cloud radii ($r_i/r_o$) at 0.01.
 The opening angle as discussed above controls the behaviour of the silicate feature
in the face-on case. The ratio of inner to outer disc radii
controls the width of the spectral energy distribution
but also the silicate feature strength because, for the same
equatorial optical depth, a more extended disc is more optically
thin in lines of sight perpendicular to the plane of the disc
than a more compact disc. The reason for choosing this particular
model is that it gives the strongest silicate emission feature
when the torus is viewed face-on and can therefore match the
silicate strength of the quasars in the sample. The spectral energy distribution of
the torus is computed for 74 inclinations which are equally
spaced in the interval 0 to 90$^o$.

\subsection{Starburst models}

Efstathiou, Rowan-Robinson \& Siebenmorgen (2000; hereafter ERRS00) developed
a starburst model that has three main features. The first
feature of the model is that it incorporates the stellar population synthesis
model of Bruzual \& Charlot (1993) that gives the spectrum of the stars
as a function of their age. We use the  table that
assumes a Salpeter IMF and stellar masses in the range
0.1-125 $M_\odot$. The second important feature of the model
is that we carry out detailed radiative transfer that takes
into account multiple scattering from grains and incorporates
a dust model that includes small transiently heated grains
and PAH molecules as well as large classical grains. The calculation
of the emission of the small grains and PAHs is according to
the method of Siebenmorgen \& Kr\"{u}gel (1992). The third feature
of the model is that it incorporates a simple model for the evolution
of the giant molecular clouds that constitute the starburst once
a stellar cluster forms instantaneously at the center of the cloud. The most
important characteristic of this evolutionary scheme is that
by about 10 Myrs after star formation, the expansion of
the HII region leads to the formation
of a cold narrow shell of gas and dust. This naturally explains
why the mid-infrared spectra of starburst galaxies are 
dominated by the PAH emission and not by the emission of
hot dust. The predicted spectra of the molecular clouds
therefore shift to longer wavelengths with age and show
stronger PAH equivalent widths. Another effect of ageing
is that the clouds get more otically thin and 
therefore the silicate features get shallower.
The sequence of molecular clouds at different ages can
be convolved with a star formation history to give the
spectrum of a starburst at different ages.

 ERRS00 showed that the IRAS colours of starburst galaxies
can be modeled with starbursts whose star formation rate
decays exponentially with an e-folding time $\tau $ of 20Myrs.
The choice of the other model parameters assumed by ERRS00
(average density $n_{av}$, star formation efficiency $\eta$
and giant molecular cloud mass $M_{GMC}$),
 was discussed by ERRS00 (section 3) and was found
to be in agreement with other indicators for the starburst galaxy M82.
This choice of parameters gave rise to an initial optical depth
of the molecular clouds in the V band $\tau_V$  of 50. Rowan-Robinson \& Efstathiou
(1993) showed that ultraluminous infrared galaxies such as Arp220 are
up to a factor of four more optically thick than starbursts
like M82 and NGC1068. Rather than computing a four-parameter grid 
of models, which would be impractical, we fixed $\tau$ at 20Myrs
and varied  $n_{av}$,  $\eta$ and  $M_{GMC}$ to give four
discrete values of $\tau_V$ (50, 100, 150 and 200) 
that span the range suggested by observations.
In other words we treat $\tau_V$ as an effective free parameter of
the model. The spectra are computed for 11 different ages
ranging from 0 to 72Myrs. This sequence of models, in combination
with the AGN torus models discussed above, has been shown
by Farrah et al (2003) to be in good agreement with the spectral
energy distributions of 41 ultraluminous infrared galaxies. 

\subsection{Cirrus models}
 
Efstathiou \& Rowan-Robinson (2003; hereafter ERR03) presented an extension
of the ERRS00 model that allowed the calculation of cirrus
models as well as a combination of starburst and cirrus. 
This model is much simpler than the GRASIL model developed by
Silva et al (1998) or the model of Piovan et al (2006) but
as it has been shown by ERRO3 the calculated spectra are in
good agreement with the spectral energy distributions of
local galaxies dominated by cirrus. The first step in this model
involves the specification of the spectrum
of starlight for which we have two options. We either use 
the Bruzual \& Charlot table and an assumed star formation
history or we use the spectrum of the interstellar radiation
field in the solar neighbourhood (ISRF; Mathis et al 1983).
In the first option we assume (as in the case of the starburst model)
that the star formation rate
declines exponentially with an e-folding time $\tau $.
For the emission of stars that formed in the last $t_m$ years
we use the spectrum computed by the starburst code. We then
scale the spectrum of starlight by varying the parameter $\psi$
which is the ratio of the bolometric intensity of starlight
to the bolometric intensity of the local ISRF. Finally we illuminate
the dust with the starlight and calculate its infrared spectrum.
The code also provides the option to attenuate the starlight,
by assuming a visual extinction $A_V$, and 
calculate the spectrum from the UV to the millimeter by
self-consistently reprocessing the energy absorbed in the
optical/UV to the infrared.

\section{The Spoon diagram}

We now test whether our suite of models, tuned to fit the broad-band SEDs from 1 $\mu$m to 1 mm,
can explain the distribution of galaxies in the diagnostic diagram developed by Spoon et al (2007),
strenght of 9.7 $\mu$m  silicate feature versus equivalent width of 6.2 $\mu$m PAH feature.
A difficulty in defining the strengths of features in the mid infrared is defining the underlying continuum.
The spectra are almost always very strongly modified both by the broad PAH emission features and by
the broad silicate (generally absorption) features.  Spoon et al (2007) used several approaches to define the 
continuum, usually employing spline fits to the emission at three selected wavelengths.  The resulting
continua are plausible but somewhat arbitrary.  Here we adopt a more uniform approach, although it is
no less vulnerable to contributions from unaccounted for absorption or emission features.
In the region of the 6.2 $\mu$m feature we use a log-log interpolation between 5.85 and 6.6 $\mu$m and in 
the region of the 9.7 $\mu$m feature we use log-log interpolation between 8.2 and 12.0 $\mu$m, to define 
the underlying continuum.  We are therefore assuming the latter has a power-law form between these
selected end-points.

Figure 1 shows the Spoon diagram for a selection of our starburst models (Efstathiou et al 2000), compared 
with the Spoon et al (2007) data for 180 starburst galaxies or AGN, showing the variation of just two
model parameters, the age of the starburst, and the initial optical depth to the centre of a GMC. 

For AGN dust tori we indicate in Fig 1 the range of 9.7 $\mu$m feature strengths as a function of
viewing angle for one of these AGN dust torus models, a tapered disk with opening angle 60$^o$ 
(see above for other parameters).  
The 6.2 $\mu$m EW is zero for these dust torus models, since we assume that in AGN environments the
intensity of UV radiation is too high for the survival of very small grains or large molecules.

Mixture lines are shown between a type 1 AGN dust torus (face-on viewing) and two of our starburst models,
those which have been used by Rowan-Robinson (2001) and Rowan-Robinson et al (2005,2008)
to approximate to an M82 starburst and an Arp 220 (higher optical depth) starburst, to a young starburst
(t = 0, $A_V$ = 150), and to an old starburst (t = 72 Myrs, $A_V$ = 50).

\begin{figure*}
\epsfig{file=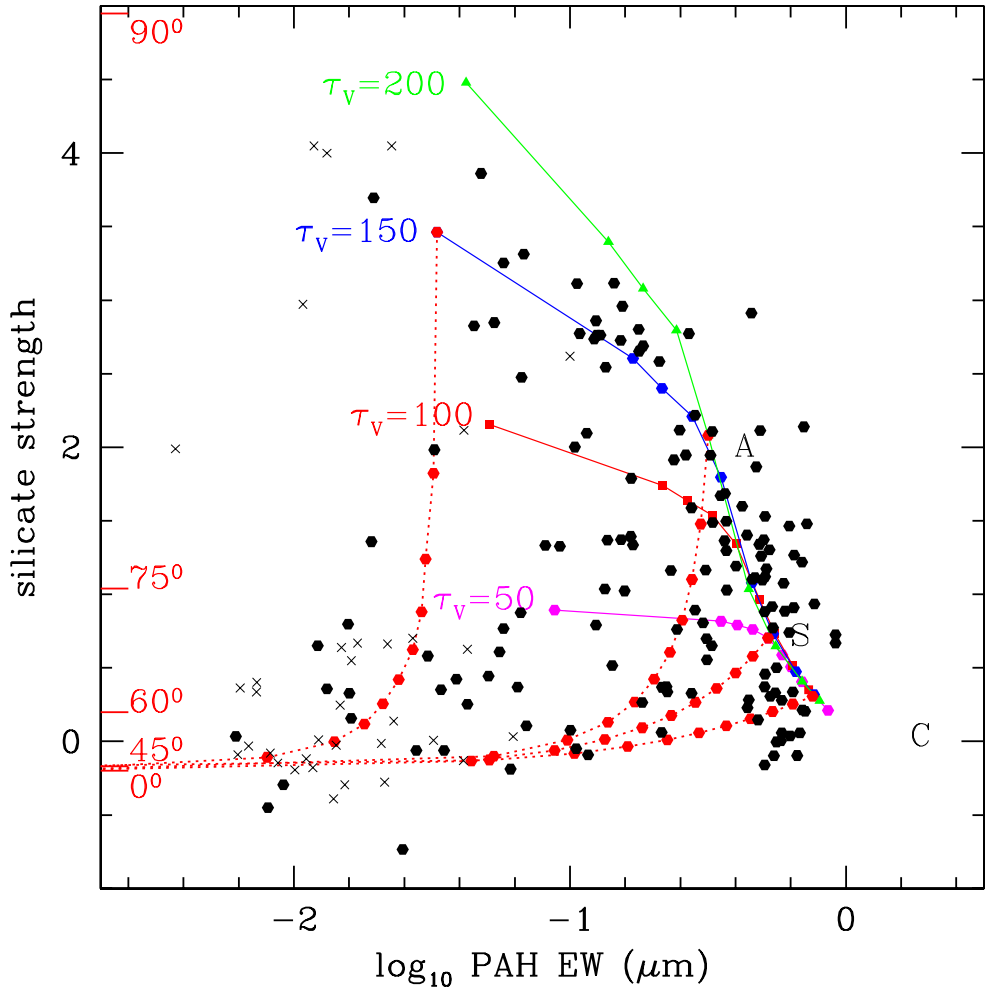,angle=0,width=14cm}
\caption{The Spoon diagram (Spoon et al 2007): optical depth of 9.7 $\mu$m silicate feature versus
equivalent width of 6.2 $\mu$m PAH feature.  Black filled circles are data points from Spoon et al
(2007), crosses denote upper limits to 6.2 $\mu$m EW.  Green, blue, red and magenta solid curves are 
sequences of starburst models with ages 0 (top)-6.6-10-16-26-37-45-56-72 (bottom)
 Myrs and $A_V$ = 200, 150, 100 and 50, respectively, from Efstathiou et al (2000).  Labels A, S and C
denote 'Arp 220', and 'M82' starbursts, and 'cirrus' (quiescent) galaxy templates used by Rowan-Robinson (2000).
Red dotted curves are mixture lines between a face-on AGN dust torus and four starbursts with (t (Myrs), $A_V$) =
(0, 150), (26, 200), (26, 50) and (72, 50). 
Red bars on the left axis denote the 9.7  $\mu$m silicate optical depth for AGN dust torus model (opening angle 60$^0$) 
viewed from inclinations 0, 45, 60, 75, 90$^0$..
}
\end{figure*}

The models bracket the observed distribution remarkably well.  Young starbursts are well differentiated by the 
 9.7 $\mu$m feature strength for different $A_V$ but at later times the sequences for different $A_V$ 
converge, ie there is aliasing between starburst age and the optical depth.  There is also aliasing between 
starbursts of young age and mixtures between AGN dust tori and older starbursts.  This diagram on its own is not capable
of resolving these aliases.   The full IRS spectrum does offer a stronger handle on model parameters, especially when
combined with photometry at longer wavelenths.

To illustrate the latter we also show, in Fig 2, the 100-60-25 (LH) and 850-100-60 (RH) $\mu$m colour-colour diagrams, 
for our starburst model sequences and for a series of cirrus models in which optically thin dust is illuminated by
light with the spectrum of the local interstellar radiation field, with intensity $\psi$ = 1, 2, 5, 10, where $\psi$ 
denotes the ratio of the intensity of the radiation field to that in the solar neighbourhood.
 The 100-60-25 $\mu$m data are from the Imperial IRAS FSC redshift (IIFSCz) catalogue 
of Wang and Rowan-Robinson (2009), with the restriction z$<$0.3.  The 850-100-60 $\mu$m data are from Dunne et al (2000),
Rigopoulou et al (1996), Fox (2002), Clements et al (2009, in preparation) and from a literature compilation by Chanial (2009, 
in preparation), with the restriction that z $<$ 0.2.  The 100-60-25 $\mu$m diagram is sensitive to both the age of the starburst and
the optical depth.  Objects with higher values of S100/S60 require a quiescent ('cirrus') contribution.
A complication of these far infrared colour-colour diagrams as diagnostics is the effect of redshift and there is
a strong benefit in using rest-frame features as in Fig 1.

While the main ridge of objects in the Spoon diagram (Fig 1) could be modeled as a single sequence of
starbursts of different ages, with $A_V \sim 150$, It is clear from the 100-60-25 $\mu$m diagram (Fig 2L)
that the full range of $A_V$ from 50 to 200 is needed.  The broad-band colour-colour diagrams (Figs 2L and 2R)
both require that the starbursts models be mixed with an optically thin 'cirrus' contribution, with $\psi$ in the range
2-10.   Some of the outliers in the  850-100-60 $\mu$m colour-colour diagram (Fig 2R) may be due to
inaccurate 850 $\mu$m integrated fluxes, but there is a hint that some galaxies may contain more cold dust
than predicted in these models.
Finally both the Spoon diagram, the 100-60-25 $\mu$m diagram (Fig 2L), and individual SEDs demonstrate the need in many cases
for an additional component due to an AGN dust torus.  Thus the broad set of components (starbursts, cirrus, AGN dust torus) 
originally introduced by Rowan-Robinson and Crawford (1989) to account for the IRAS 12-100 $\mu$m colours still hold, 
and encompass what we see in infrared galaxy SEDs, albeit with considerable refinement.  

\begin{figure*}
\epsfig{file=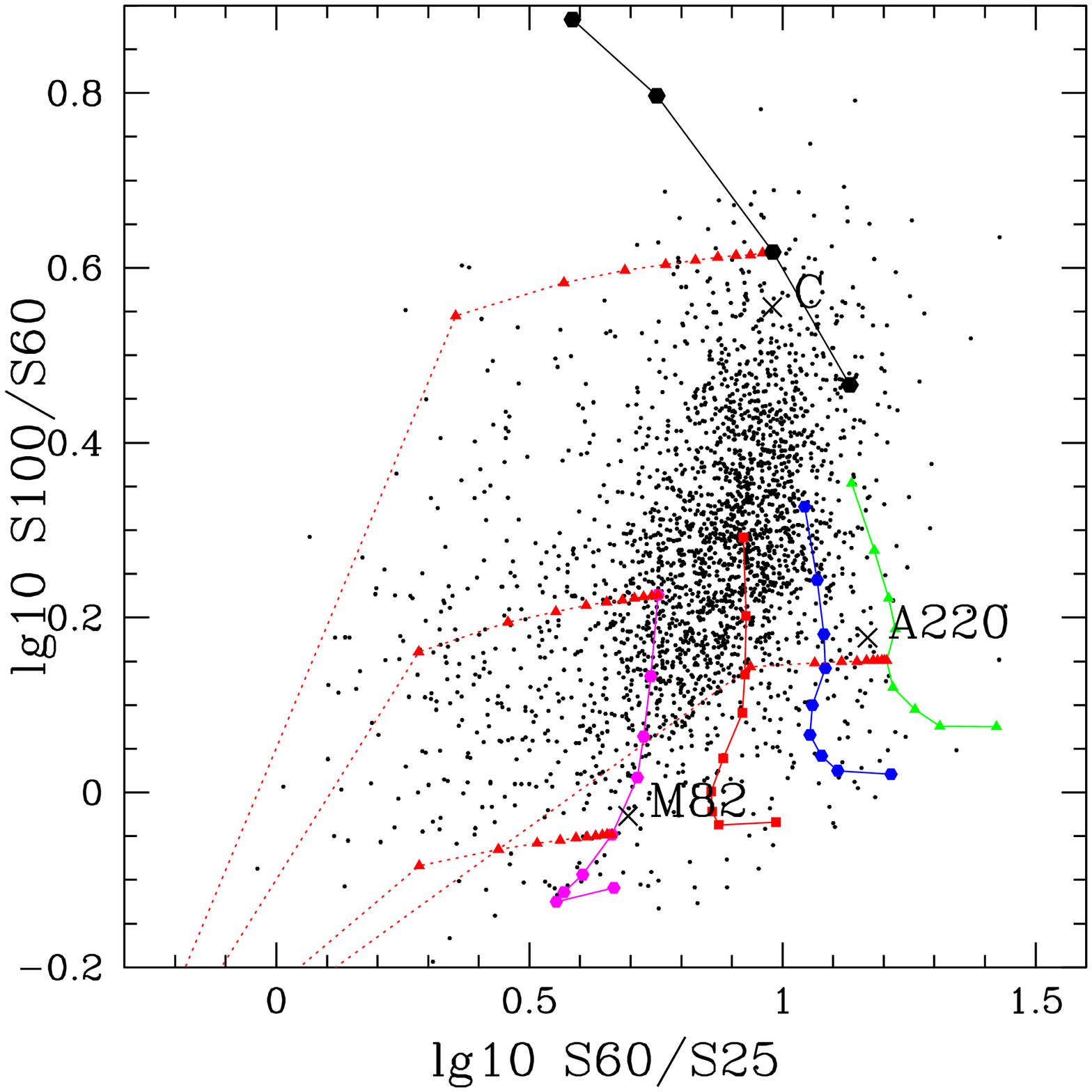,angle=0,width=7cm}
\epsfig{file=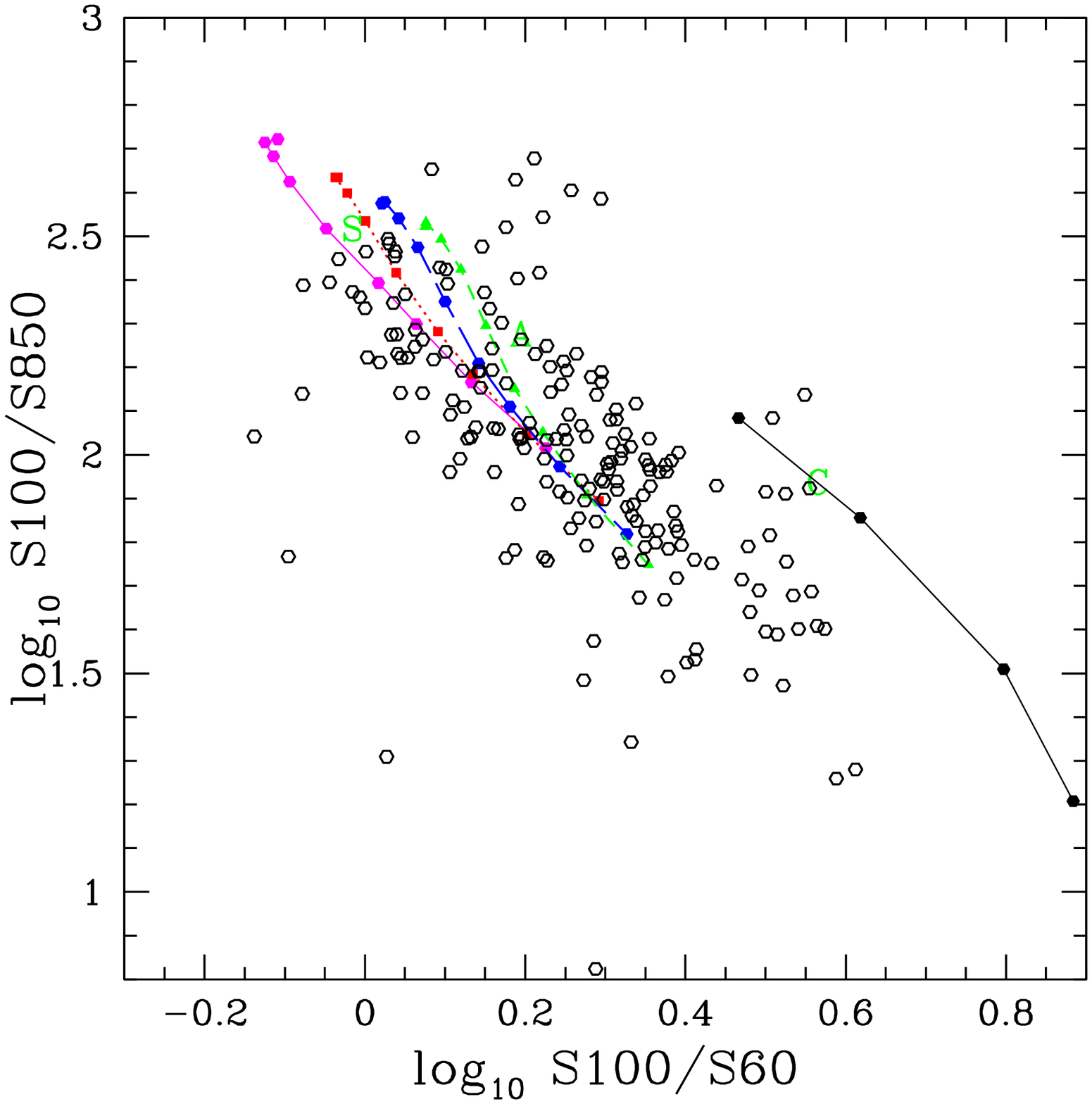,angle=0,width=7cm}
\caption{Far infrared colour-colour diagrams.  LH: 100-60-25 $\mu$m, data from IIFSCz catalogue (Wang and
Rowan-Robinson 2009). RH 850-100-60 $\mu$m, data from literature compilation by Chanial (2009, in preparation).
Starburst and AGN dust torus mixture models as in Fig 1.  Youngest starburst models have lowest values of S(100)$/$S(60).   
Black filled circles and solid curve: sequence of cirrus models with
 $\psi$ = 1, 2, 5, 10 (lowest value of S(100)$/$S(60)). 
.}
\end{figure*}

\section{Detailed fits to SEDs of average spectral classes from Spoon et al 2007}

Spoon et al (2007) divide their diagnostic diagram into 8 regions and generate a mean IRS spectrum
for the galaxies in each region.  In Figs 3-6 we show some of our model fits for these mean 
IRS spectra.  Note  that our models do not incorporate the 12.8 $\mu$mu NeII line.  Fits could probably be improved
with a finer grid of models.   In Figure 5R we have included a cirrus model to show how similar this is to 
an old starburst at 5-40 $\mu$m.

Such average SEDs give us some idea of the minimal set of templates needed to characterize infrared
galaxy spectra.  But because of the strong evolution of individual SEDs as a starburst ages and the possible aliasing 
between starburst age, initial optical depth, and possible presence of an AGN dust torus, galaxies at very different 
stages of their evolution may be being combined together in these averages.  The 1A average spectrum (Fig 6)
certainly looks like an average together of AGN dust tori viewed at very different inclinations.  In future work we plan to
model a substantial sample of individual galaxy SEDs where IRS data are available.

\begin{figure*}
\epsfig{file=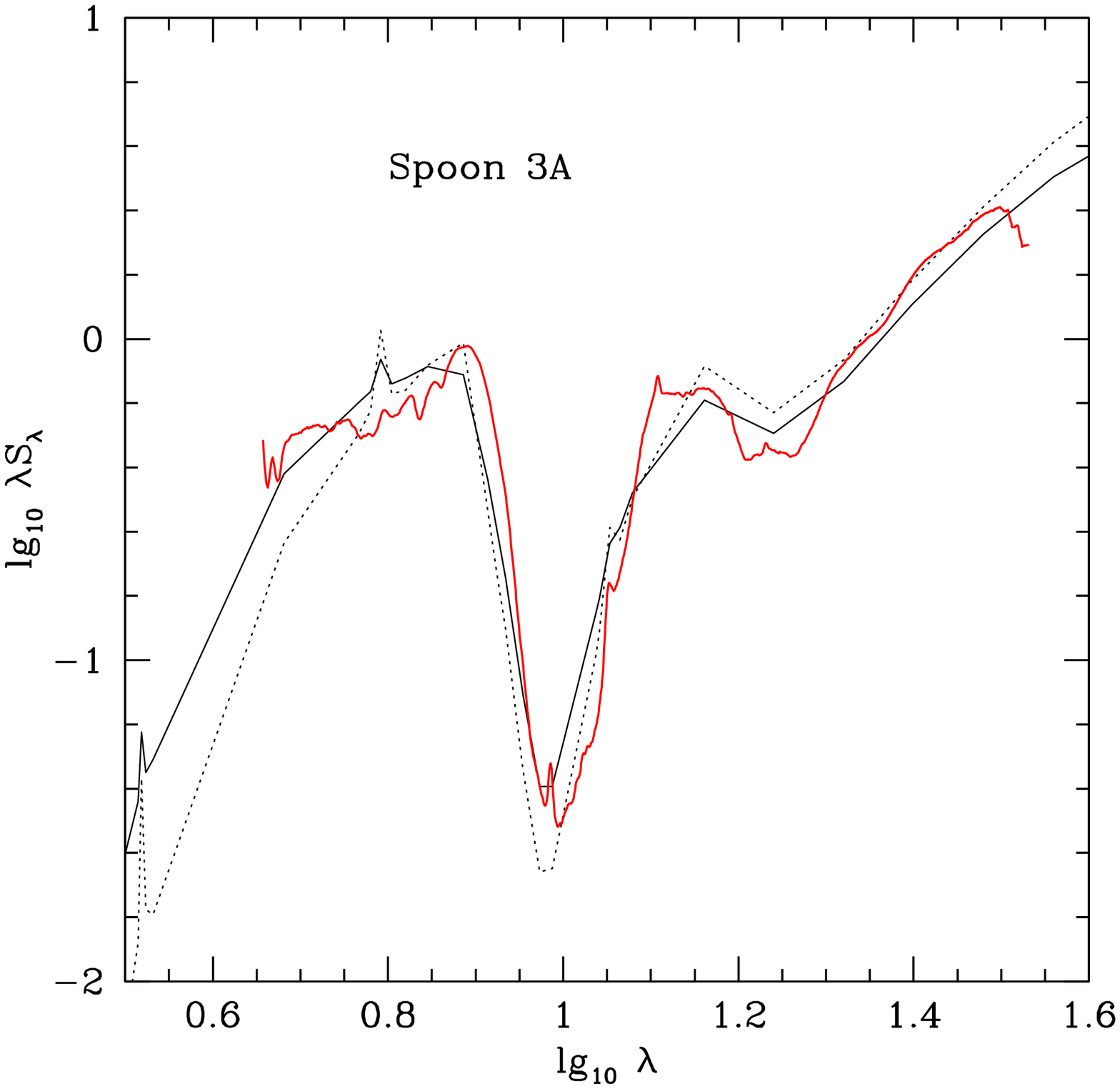 ,angle=0,width=7cm}
\epsfig{file=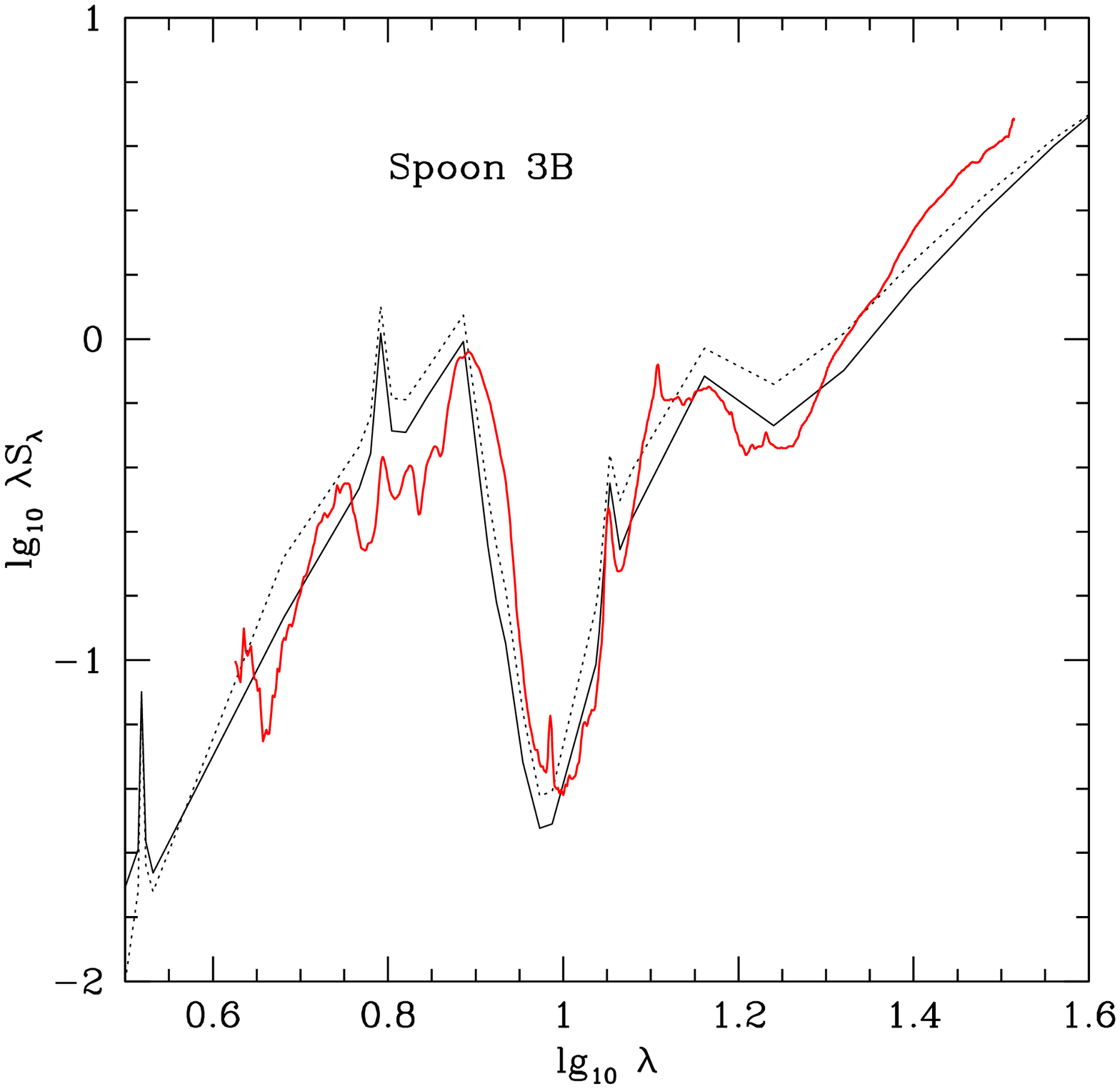 ,angle=0,width=7cm}
\caption{LH: SED fits to Spoon et al (2006) mean spectral class 3A.  Solid curve: starburst t=0 Myrs, $A_V$=100,
dotted curve: starburst t=6.6, $A_V$=150.
 RH: mean spectral class 3B.  Solid curve: starburst t=26 Myrs, $A_V$=200, dotted curve: starburst t=16, $A_V$=150.}
\end{figure*}

\begin{figure*}
\epsfig{file=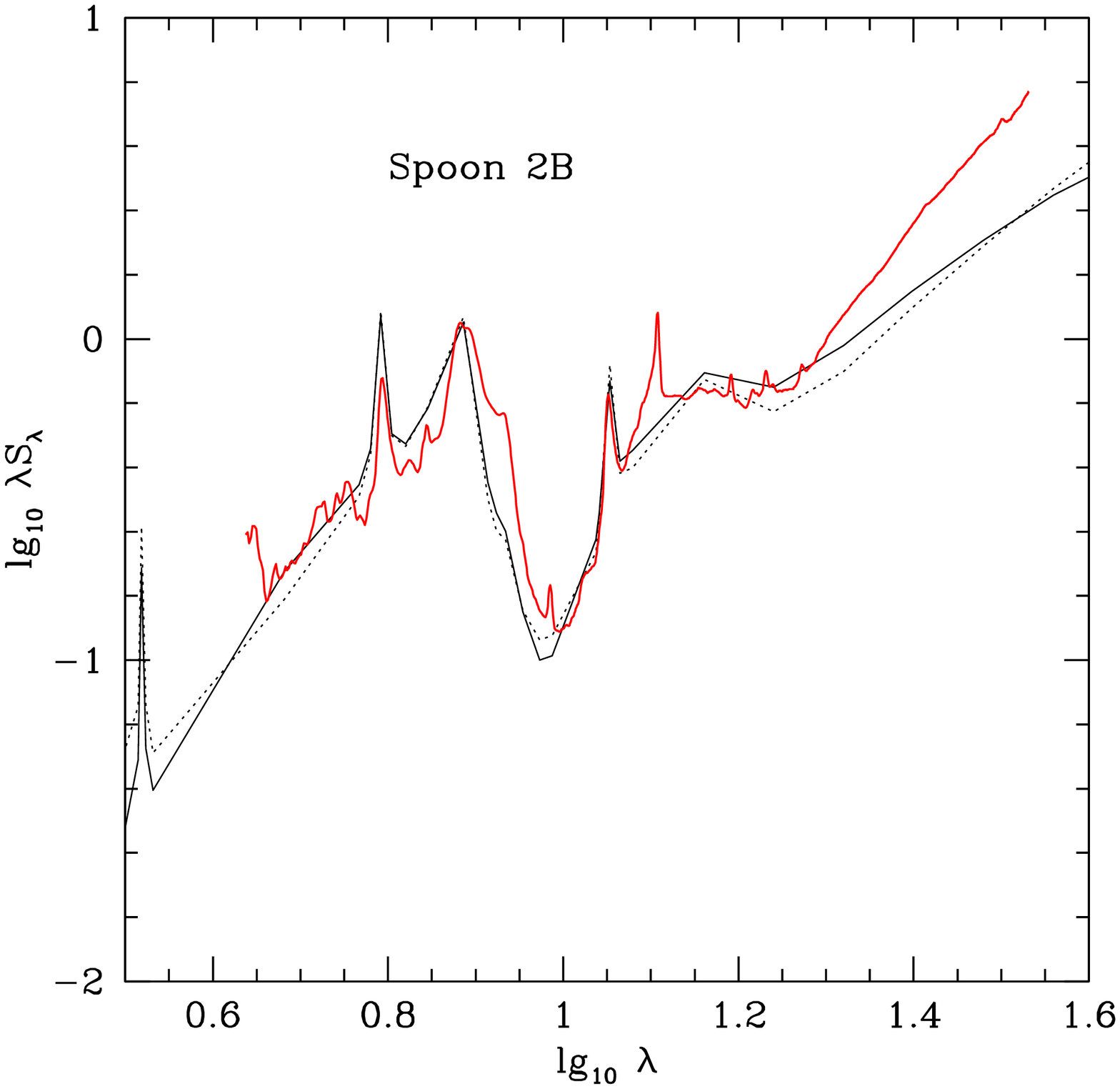 ,angle=0,width=7cm}
\epsfig{file=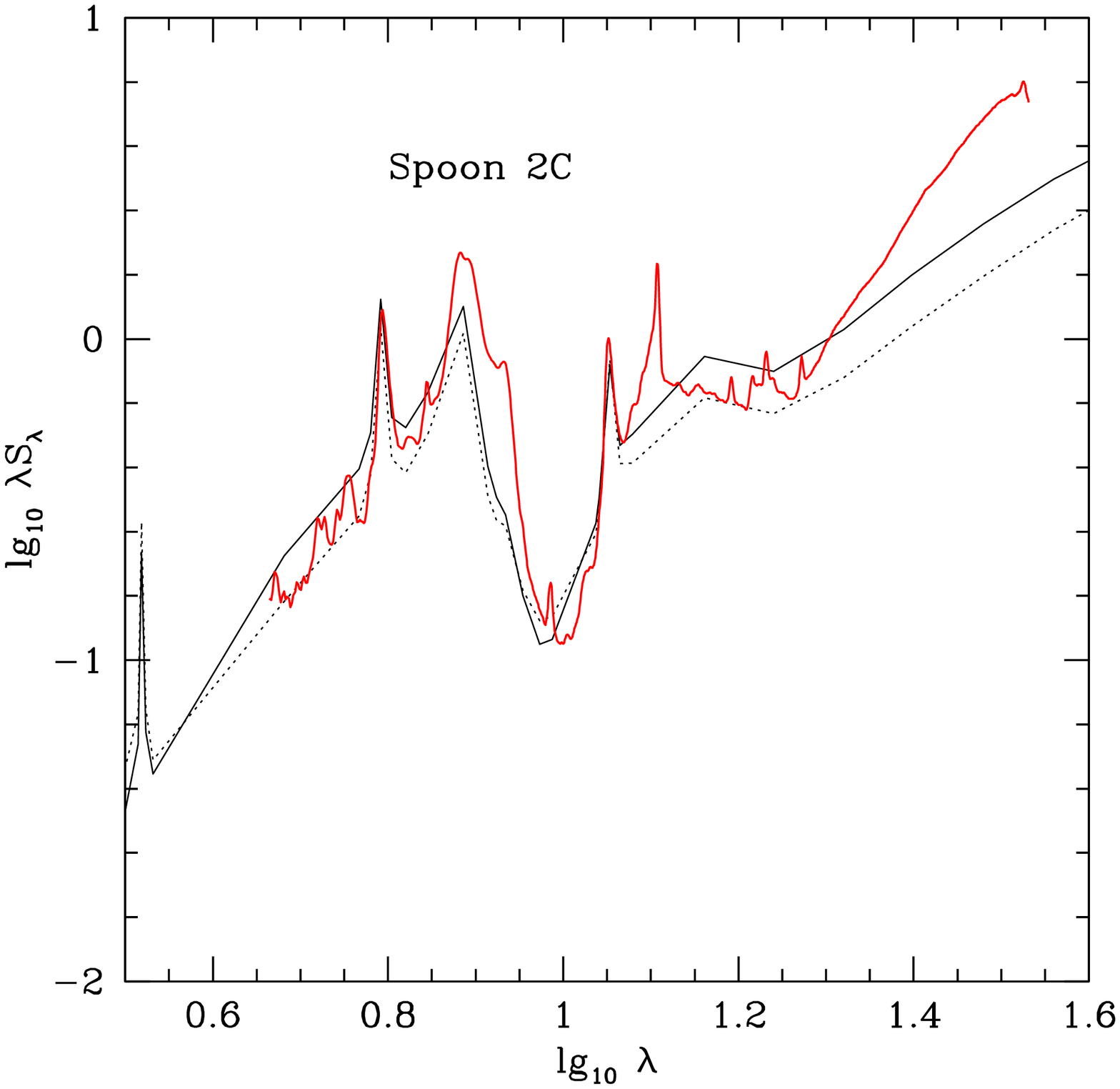 ,angle=0,width=7cm}
\caption{LH: SED fits to Spoon et al (2006) mean spectral class 2B.  Solid curve: starburst t=26 Myrs, $A_V$=100,
 dotted curve: starburst t=37 Myrs, $A_V$=150.
 RH: mean spectral class 2C.  Solid curve: starburst t=26 Myrs, $A_V$=100, dotted curve: starburst t=37 Myrs, $A_V$=100.}
\end{figure*}

\begin{figure*}
\epsfig{file=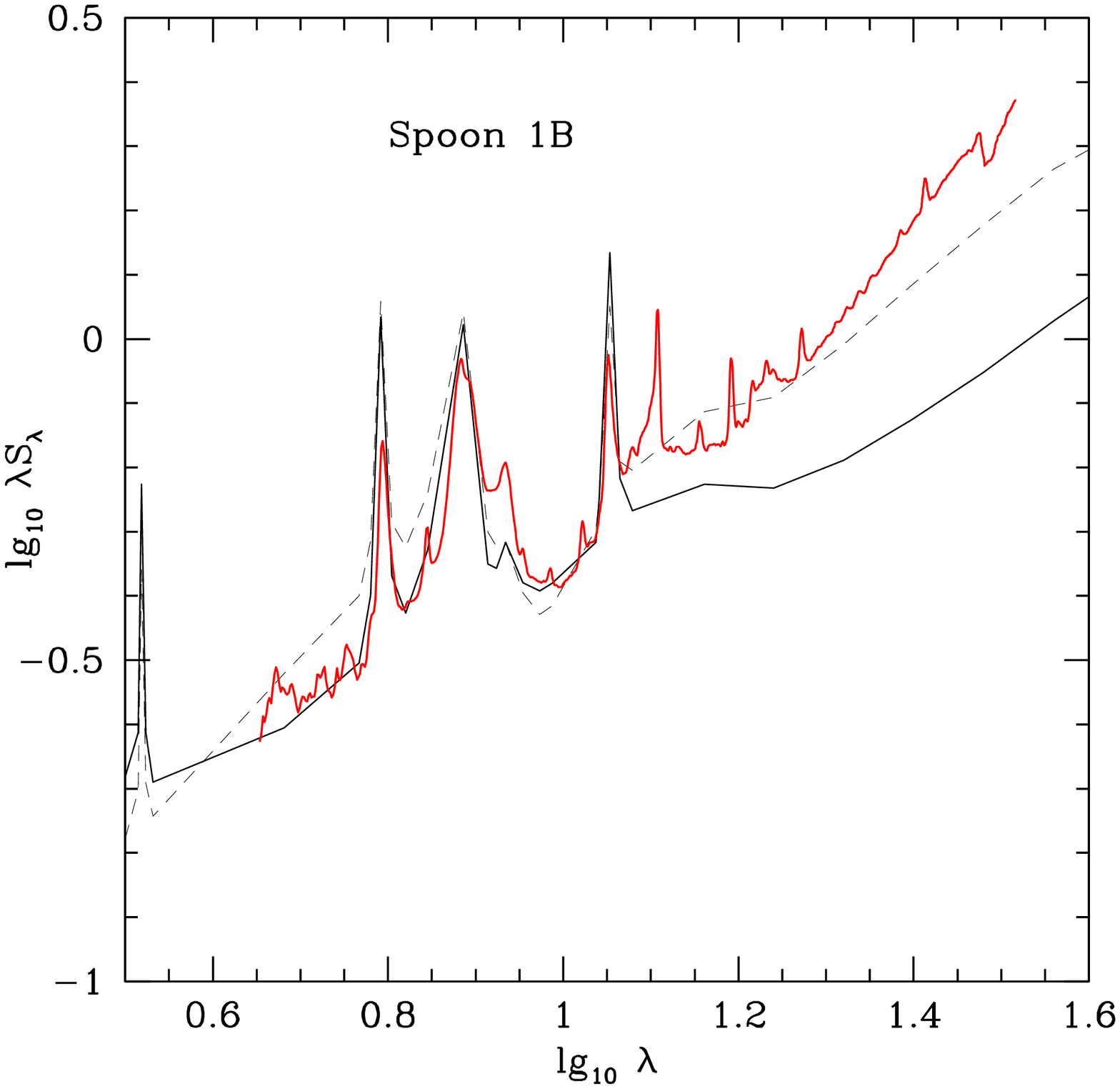 ,angle=0,width=7cm}
\epsfig{file=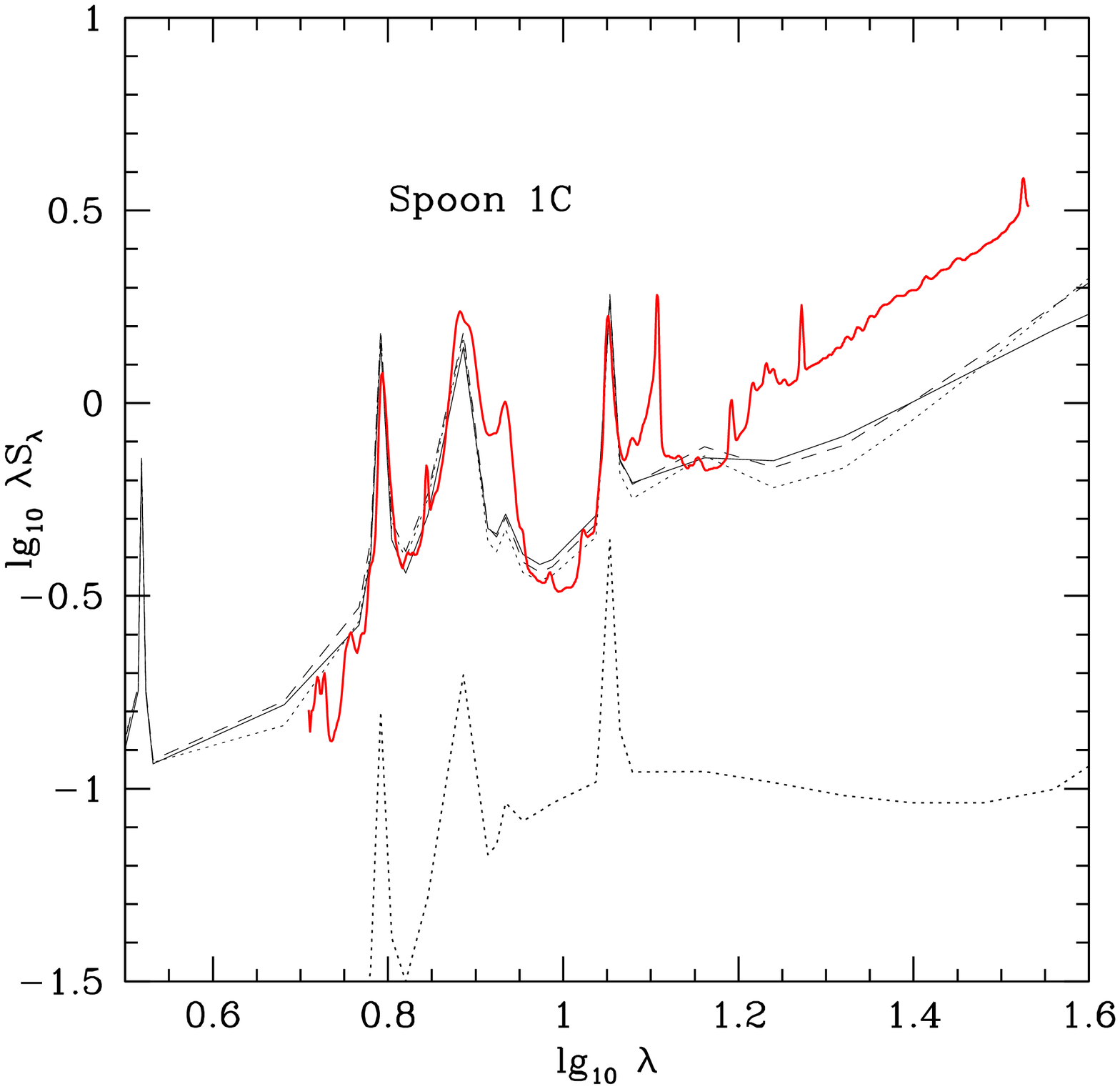 ,angle=0,width=7cm}
\caption{LH: SED fits to Spoon et al (2006) mean spectral classes 1B. Solid curve: mixture of starburst t=72 Myrs, $A_V$=50
and face-on AGN dust torus, broken curve: mixture of starburst t=26Myrs, $A_V$=50
and face-on AGN dust torus.
RH: mean spectral class 1C. Solid curve: starburst t=72 Myrs, $A_V$=50,
broken curve: starburst t=72 Myrs, $A_V$=100, dotted curve: starburst t=72 Myrs, $A_V$=150. Displaced dotted curve: cirrus model with $\psi$ = 5.}
\end{figure*}

\begin{figure*}
\epsfig{file=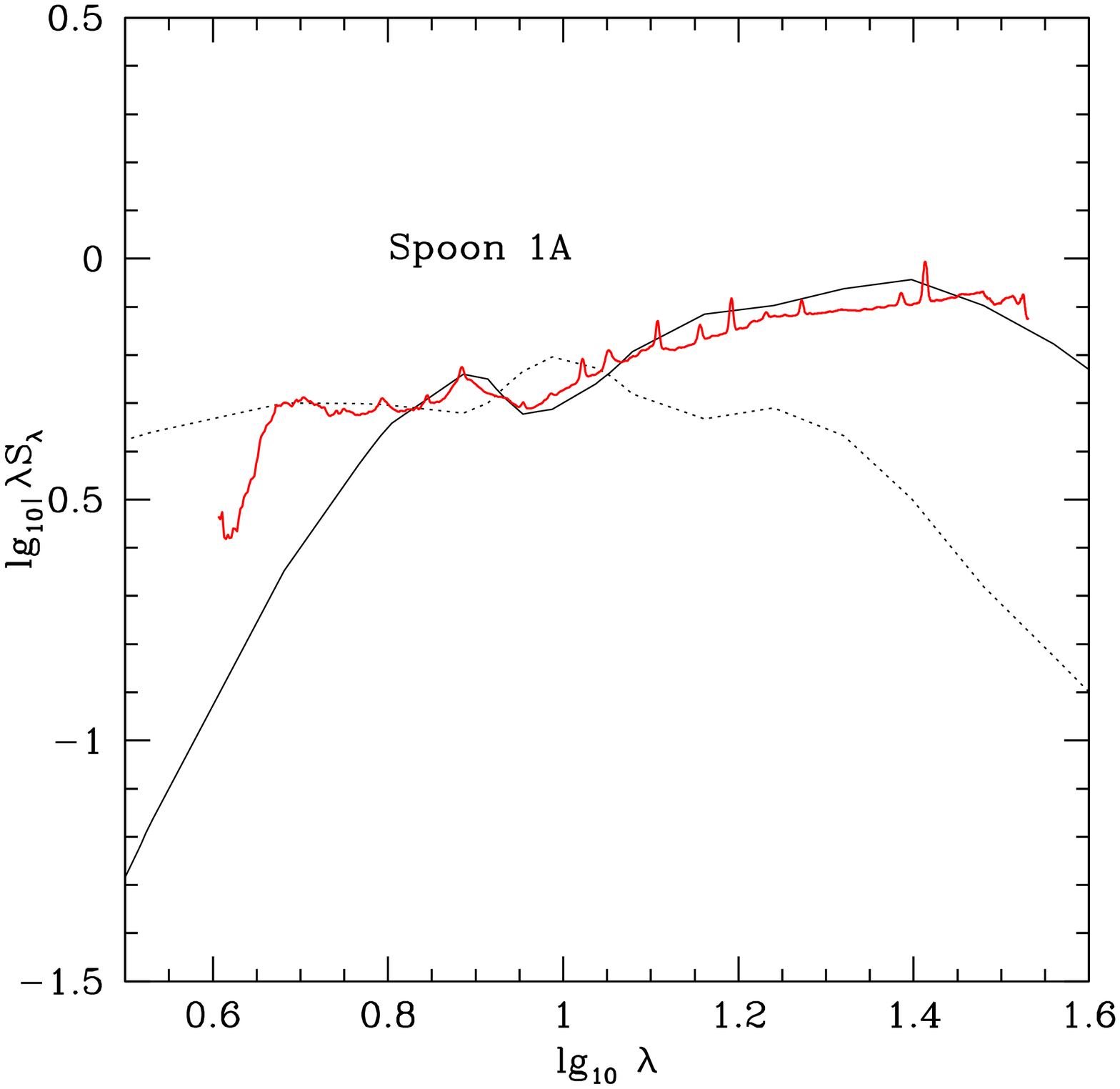 ,angle=0,width=7cm}
\caption{SED fits to Spoon et al (2006) mean spectral classes 1A. Solid curve: AGN dust torus, inclination 77$^0$.
Dotted curve: AGN dust torus, face-on.
}
\end{figure*}

\section{Discussion and conclusions}
Our models for starbursts, quiescent star-forming galaxies and for AGN dust tori are able to model the full range
of IRS spectra measured with {\it Spitzer}.  The Spoon et al (2007) diagnostic diagram gives a good indication of the age
and optical depth of a starburst, and of the contribution of an AGN dust torus.  There are good possibilites of
determining the age of a starburst from its spectrum and this has important implications for understanding the
evolution of starburst and quasar activity in galaxies.  However there is aliasing
between age and optical depth at later times in the evolution of a starburst, and between age and the presence of
an AGN dust torus. 

The use of a small number of templates to model far infrared and submillimetre SEDs (Rowan-Robinson 2000)
has obvious limitations, given the large number of potential model parameters.  A significant improvement to
future analyses of this type might be to add an additional template corresponding to a young 
starburst,  as in the mixture lines shown in Fig 1. Although these represent a small fraction of the infrared
galaxy population ($<10\%$) they are of great interest for evolutionary studies. 
An old starburst, which might also be desirable to bracket the
full range of observed starburst, differs only slightly from a quiescent (cirrus) galaxy, at least in the mid infrared (see Fig 5R). 


\section{Acknowledgements}

We thank Henrik Spoon for supplying the data for the Spoon et al (2007) diagnostic diagram and the
mean IRS spectra for their spectral classes.


\end{document}